\documentclass[aps,prb,preprint,groupedaddress]{revtex4-1}

\usepackage{graphicx,amsmath}

\begin{document}

%\preprint{}

\title{Magnetic structure of noncentrosymmetric perovskites PbVO$_3$ and BiCoO$_3$}


\author{I. V. Solovyev}
\email{SOLOVYEV.Igor@nims.go.jp}
%\thanks{}
\affiliation{
National Institute for Materials Science, 1-2-1 Sengen, Tsukuba,
Ibaraki 305-0047, Japan}


\date{\today}

\begin{abstract}
It is well known that if a crystal structure has no inversion symmetry,
it may allow for Dzyaloshinskii-Moriya magnetic interactions,
operating between different crystallographic unit cells,
which in turn should lead to the formation of long-periodic
spin-spiral structures. Such a behavior is anticipated for two
simple perovskites PbVO$_3$ and BiCoO$_3$, crystallizing in the
noncentrosymmetric tetragonal $P4mm$ structure.
Nevertheless, we argue that in reality PbVO$_3$ and BiCoO$_3$ should behave
very differently. Due to the fundamental Kramers degeneracy for the
odd-electron systems, PbVO$_3$ has no single-ion anisotropy. Therefore, the ground state of
PbVO$_3$ will be indeed the spin spiral with the period of about one hundred unit cells.
However, the even-electron BiCoO$_3$ has a large single-ion anisotropy, which locks
this system in the collinear easy-axis C-type antiferromagnetic ground state. Our theoretical
analysis is based on the low-energy model, derived from the first-principles electronic
structure calculations.
\end{abstract}

%\keywords{}

\maketitle

\section{\label{Intro} Introduction}

  Magnetic materials, crystallizing in the noncentrosymmetric structure, have attracted a
great deal of attention. The lack of the inversion symmetry gives rise to the ferroelectric activity.
If the latter property is combined with the magnetism, the system becomes multiferroic,
which has many merits for the next generation of electronic devices:
for example, one can control the magnetization by applying the electric field
and vice versa. The canonical
example of such materials is BiFeO$_3$, which possesses simultaneously high magnetic transition
temperature (about $640$ K) and high ferroelectric Curie temperature (about $1090$ K).\cite{Lebeugle}

\begin{figure}[h!]
\begin{center}
\includegraphics[height=7.0cm]{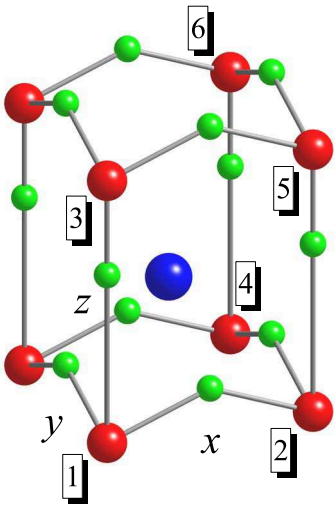}
\end{center}
\caption{\label{fig.structure} (Color online) Fragment of the crystal structure
of BiCoO$_3$. The Bi atoms are indicated
by the big blue (dark) spheres, the Co atoms are indicated
by the medium red (dark grey) spheres, and the oxygen atoms are indicated
by the small green (light grey) spheres. }
\end{figure}

  Recently fabricated PbVO$_3$ and BiCoO$_3$ belong to the same category.
They crystallize in the noncentrosymmetric tetragonal $P4mm$ structure (Fig.~\ref{fig.structure}).\cite{Belik05,Belik06}
BiCoO$_3$ is an antiferromagnetic (AFM) insulator of the C-type with the
N\'eel temperature of about $470$ K.\cite{Belik06} The experimental information about
PbVO$_3$ is rather controversial.\cite{Shpanchenko,Tsirlin}
The magnetic susceptibility has a broad maximum
around $200$, which might be the sign of the antiferromagnetism.
On the other hand, no long-range magnetic order was found down
to $1.8$ K in the neutron diffraction experiments. However, the analysis of the neutron data
depends on the model of the magnetic structure, which is typically assumed in the process of
interpretation. Finally, the experimental studies of PbVO$_3$ were hampered by
possible defects in the sample.\cite{Tsirlin}

  According to first-principles electronic structure calculations, both PbVO$_3$ and
BiCoO$_3$ are expected to have the C-type AFM ground state, although in PbVO$_3$
it is nearly degenerate with the G-type AFM state.\cite{Uratani,Singh}
Giant electric polarization (more than $150$ $\mu$C/cm$^2$) was predicted
theoretically both for PbVO$_3$ and BiCoO$_3$,\cite{Uratani} which spurred additional interest to these
systems.

  Nevertheless, the violation of the inversion symmetry gives rise to a number of interesting effects,
which are not currently accessible by the
first-principles electronic structure calculations, simply due to their complexity. One of them is a
complex magnetic ordering, caused by antisymmetric Dzyaloshinskii-Moriya
interactions: in the noncentrosymmetric systems, these interactions, of the relativistic origin,
can operate between different
crystallographic unit cells, thus driving the formation of
long-periodic spin-spiral superstructures.\cite{DM}
Particularly, the
idea of the spin-spiral order in various oxide materials has attracted much attention recently
in the context of their multiferroic behavior and was proposed as one of the possible origins
of such behavior.\cite{IDM}

  In this paper we will address some basic issues of the formation of the spin-spiral states
in PbVO$_3$ and BiCoO$_3$. We will argue that, despite similarities in
the lattice distortion and population of the crystal-field levels, these
two compounds will behave very differently.
Particularly, we will show that the long-periodic spin spiral is a probable
candidate for the magnetic ground state of PbVO$_3$, where due to the fundamental Kramers degeneracy,
the single-ion anisotropy does not exists. On the contrary, the spin-spiral state
in BiCoO$_3$ (for which the Kramers theorem is no longer applicable) is suppressed
by the single-ion anisotropy, which reinforces the formation of the easy-axis collinear
C-type AFM ground state. Our analysis is based on the low-energy model, derived from the
first-principles electronic structure calculations. In this sense, this is the continuation of
our previous works, devoted to `realistic modeling' of complex oxide materials
and other strongly correlated systems.\cite{review2008,BiMnO3PRB10,TbMnO3PRB11}

  The rest of the paper is organized as follows. In Sec.~\ref{Method} we briefly discuss
the construction of the low-energy model (in our case -- the multiorbital Hubbard model)
on the basis of first-principles electronic structure calculations. All model parameters
can be found in the supplemental materials.\cite{SM} Sec.~\ref{SpinModel} is devoted to
semi-quantitative analysis of the spin model, which can be derived from the
multiorbital Hubbard model. Particularly, we consider the formation of
incommensurate spin-spiral states, resulting from the competition of the isotropic exchange and
Dzyaloshinskii-Moriya interactions, and explain the main difference in the behavior of the single-ion anisotropy
in PbVO$_3$ and BiCoO$_3$. In Sec.~\ref{ElModel}, we will present results of
extensive Hartree-Fock calculations for the long-periodic spin-spiral states in the
electronic Hubbard model. Finally, in Sec.~\ref{Summary}
we will briefly summarize the main results.

\section{\label{Method} Construction of the low-energy model}

  The magnetic properties of PbVO$_3$ and BiCoO$_3$ are mainly determined
by the behavior of $3d$-bands located near the Fermi level.
Therefore, our basic idea of our approach is to construct an effective low-energy model,
formulated in the Wannier-basis for the $3d$-bands, and to solve it by using
model techniques. More specifically, we adopt the form of the
multiorbital Hubbard model
on the lattice of transition-metal sites:
\begin{equation}
\hat{\cal{H}}  =  \sum_{ij} \sum_{\alpha \alpha'}
t_{ij}^{\alpha \alpha'}\hat{c}^\dagger_{i\alpha}
\hat{c}^{\phantom{\dagger}}_{j\alpha'} +
  \frac{1}{2}
\sum_{i}  \sum_{ \{ \alpha \} }
U_{\alpha \alpha' \alpha'' \alpha'''} \hat{c}^\dagger_{i\alpha} \hat{c}^\dagger_{i\alpha''}
\hat{c}^{\phantom{\dagger}}_{i\alpha'}
\hat{c}^{\phantom{\dagger}}_{i\alpha'''}.
\label{eqn.ManyBodyH}
\end{equation}
where we use the shorthand notations,
according to which each Greek symbol stand for the combination of spin
($s$$=$ $\uparrow$ or $\downarrow$) and orbital
($m$$=$ $xy$, $yz$, $3z^2$$-$$r^2$, $zx$, or $x^2$$-$$y^2$) indices.
All parameters of the model Hamiltonian can be derived in an \textit{ab initio} fashion,
on the basis of first-principles electronic structure calculations.
For instance, the one-electron part $t_{ij}^{\alpha \alpha'}$ was
obtained by using the downfolding procedure, and the Coulomb (and exchange) interactions
$U_{\alpha \alpha' \alpha'' \alpha'''}$ -- by combining
the constrained density-functional theory (DFT) with
the random-phase approximation (RPA).
The method was discussed in the literature, and for details the reader is
referred to Ref.~\onlinecite{review2008}. Resent applications to multiferroic
compounds can be found in Refs.~\onlinecite{BiMnO3PRB10} and \onlinecite{TbMnO3PRB11}.
In all calculations we use experimental parameters of the crystal
structure, reported in Refs.~\onlinecite{Belik05,Belik06}.

  Without spin-orbit interaction, $t_{ij}^{\alpha \alpha'}$
is diagonal with respect to the spin indices
$t^{\alpha \alpha'}_{ij} \equiv t^{m m'}_{ij} \delta_{s s'}$.
The site-diagonal part of $\hat{t}_{ij} = \| t^{m m'}_{ij} \|$
describes the crystal-field effects, while the off-diagonal part
stands for transfer integrals.

\begin{figure}[h!]
\begin{center}
\includegraphics[height=7.0cm]{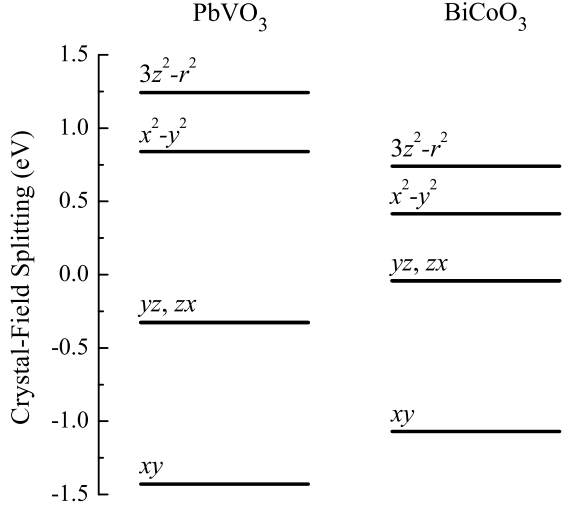}
\end{center}
\caption{\label{fig.CF} Scheme of the crystal-field splitting in PbVO$_3$ (left) and BiCoO$_3$ (right). }
\end{figure}

  The crystal field stabilizes the $xy$ orbitals (Fig.~\ref{fig.CF}).
The splitting between $xy$- and the following after them  $yz$- and $zx$-orbitals is about 1 eV,
both for PbVO$_3$ and BiCoO$_3$.
The $x^2$$-$$y^2$ and $3z^2$$-$$r^2$ orbitals lie in the higher-energy region (and are substantially higher
for PbVO$_3$, in comparison with BiCoO$_3$).
Thus, from the viewpoint of the crystal-field splitting, PbVO$_3$ and BiCoO$_3$ are expected to be very similar:
PbVO$_3$ has only one $d$-electron, which will occupy the $xy$ orbital. Amount six $d$-electrons of BiCoO$_3$,
the one for the minority-spin states will also occupy the $xy$ orbital and interact with the
spherical $d$-electron density of occupied majority-spin shell.

  The details of transfer integrals can be found in Ref.~\onlinecite{SM}.
One of the most interesting features in $\hat{t}_{ij}$ is the appearance of the so-called
`forbidden hoppings', for example between $3z^2$$-$$r^2$ and $zx$ orbitals
in the bond 1-2, which would not exist in the centrosymmetric structure
(see Fig.~\ref{fig.structure} for the notations of atomic sites).
These transfer integrals have the following form (in meV):
$$
\hat{t}_{12} =
\left(
\begin{array}{rrrrr}
 -173 & 38 &   0 &    0 & 0    \\
  -38 & 44 &   0 &    0 & 0    \\
    0 &  0 &  50 & -116 & 2    \\
    0 &  0 & 116 &  196 & -321 \\
    0 &  0 &   2 &  321 & -262 \\
\end{array}
\right)
$$
and
$$
\hat{t}_{12} =
\left(
\begin{array}{rrrrr}
 -56 &  6 &  0 &   0 &    0 \\
  -6 & 38 &  0 &   0 &    0 \\
   0 &  0 & 34 & -48 &   -3 \\
   0 &  0 & 48 & 231 & -228 \\
   0 &  0 & -3 & 228 & -164 \\
\end{array}
\right),
$$
for PbVO$_3$ and BiCoO$_3$, respectively, in the basis of
$xy$, $yz$, $3z^2$$-$$r^2$, $zx$, and $x^2$$-$$y^2$ orbitals.
The microscopic origin of such forbidden hoppings was considered in Ref.~\onlinecite{review2008}:
due to the parity violation, the Wannier orbital, which is formally labeled as
``$3z^2$$-$$r^2$'' has some weight of the $p_z$ orbitals and, therefore, can interact with the
$zx$ orbitals of the neighboring sites. Such hoppings
give rise to the antisymmetric part of $\hat{t}_{ij}$, which is responsible
for the appearance of Dzyaloshinskii-Moriya interactions.
Similar situation takes place in the bonds 1-4 and 1-5. On the contrary, due to
the rotational symmetry, the transfer integrals in the bond 1-3 are diagonal with respect to the
orbital indices (the actual values are
$t_{13}^{mm}$$=$ $-$$45$, $4$, $-$$237$, $4$, and $-$$21$ meV for PbVO$_3$, and
$t_{13}^{mm}$$=$ $-$$19$, $24$, $-$$42$, $24$, and $-$$34$ meV for BiCoO$_3$ -- other details
can be found in Ref.~\onlinecite{SM}).
Therefore,
the `forbidden hoppings' do not take place and
Dzyaloshinskii-Moriya interactions will vanish.

  For the relativistic spin-orbit interaction (SOI), we adopt two schemes.
In the first one, we evaluate SOI only at the transition-metal sites and add it
to the site-diagonal part of $t^{\alpha \alpha'}_{ij}$ in the form $\xi {\bf L} {\bf S}$,
where $\xi$$=$ $35$ and $81$ meV for PbVO$_3$ and BiCoO$_3$, respectively.
Thus, the effects of the SOI are expected to be larger in BiCoO$_3$:
due to large $\xi$ and smaller crystal-field splitting (Fig.~\ref{fig.CF}), which competes with SOI.
If it is not specified otherwise, we will refer to this scheme, for which most of the
calculations have been performed. Nevertheless, as a test, we use also the second scheme,
where SOI was included at all atoms on the level of the band-structure calculations and then
corresponding parameters $t^{\alpha \alpha'}_{ij}$ were derived through the downfolding procedure.
For example, this schemes takes into account the effect of large SOI at heavy atoms Pb and Bi.

  The spin-dependence of Coulomb matrix elements
has the standard form:
$U_{\alpha \alpha' \alpha'' \alpha'''}$$=U_{m m' m'' m'''} \delta_{s s'} \delta_{s'' s'''}$.
The details of $U_{m m' m'' m'''}$ can be found in Ref.~\onlinecite{SM}.
Rough idea about the strength of the matrix elements $U_{m m' m'' m'''}$
can be obtained by interpolating them in terms of three characteristic averaged parameters
$U$, $J$ and $B$, which would take place in the centrosymmetrical environment of
isolated atoms. In these notations, $U = F^0$ is the on-site Coulomb interaction,
$J = (F^2 + F^4)/14$ is the intraatomic exchange interaction, and
$B = (9F^2 - 5F^4)/441$ is the `nonsphericity', in terms of
radial Slater's integrals $F^0$, $F^2$ and $F^4$.
In the other words,
$U$ enforce the charge stability of certain atomic configurations,
while $J$ and $B$ are responsible for
the Hund rules.
The results of such interpolation are shown in Table~\ref{tab:UJB}.
\begin{table}[h!]
\caption{Averaged values of the Coulomb interaction $U$, exchange interaction $J$, and the
nonsphericity $B$, obtained from the fitting of the matrix elements
$U_{m m' m'' m'''}$. All parameters are measured in electron volt.}
\label{tab:UJB}
\begin{ruledtabular}
\begin{tabular}{cccc}
 compound  & $U$  & $J$  & $B$  \\
\hline
 PbVO$_3$  & 1.57 & 0.84 & 0.08 \\
 BiCoO$_3$ & 2.38 & 0.90 & 0.09 \\
\end{tabular}
\end{ruledtabular}
\end{table}
One can clearly see that the on-site Coulomb repulsion $U$ is strongly screened,
especially in PbVO$_3$, while other parameters are close to atomic values.
We use this
interpolation only for explanatory purposes, while
all practical calculations were performed with actual parameters $U_{m m' m'' m'''}$
reported in Ref.~\onlinecite{SM}.
The deviation of $U_{m m' m'' m'''}$ from the centrosymmetric form is quite strong.
For example, in the case of BiCoO$_3$, the diagonal matrix elements vary as $U_{m m m m}$$=$
3.84, 3.39, 2.94, 3.39, and 3.48 eV for $m$$=$ $xy$, $yz$, $3z^2$$-$$r^2$, $zx$, and $x^2$$-$$y^2$,
respectively.

  After the construction, the model (\ref{eqn.ManyBodyH}) is solved
in the Hartree-Fock (HF) approximation.\cite{review2008}

\section{\label{Results} Results and discussions}

\subsection{\label{SpinModel} Qualitative analysis based on the spin Hamiltonian}

  The existence of the spin-spiral states in noncentrosymmetric perovskites can be understood
in the framework of the spin model:\cite{Zvezdin}
\begin{equation}
\hat{\cal H}_S = - \sum_{ \langle ij \rangle } J_{ij} \boldsymbol{S}_i \boldsymbol{S}_j +
\sum_{ \langle ij \rangle } {\bf d}_{ij} [\boldsymbol{S}_i \times \boldsymbol{S}_j] +
\sum_i \boldsymbol{S}_i
\hat{\tau}_{ii} \boldsymbol{S}_i
\label{eqn:SpinModel}
\end{equation}
(where $J_{ij}$ is the isotropic exchange interaction, ${\bf d}_{ij}$ is the vector
of antisymmetric Dzyaloshinskii-Moriya interactions, and $\hat{\tau}_{ii}$ is the
single-ion anisotropy tensor),
which can be obtained by mapping the electronic model (\ref{eqn.ManyBodyH}) onto the spin one and integrating out
all degrees of freedoms but spins.
There are several ways how to do it.
One possibility is to
consider the perturbation-theory expansion with respect to the infinitesimal
spin rotations and SOI near the nonrelativistic
ground state in the Hartree-Fock approximation.\cite{PRL96}
In the following, the results of such model will be
denoted by the symbols `$inf$'. Moreover, for the $d^1$ configuration of
PbVO$_3$ one can easily consider the theory of superexchange interactions
in the second order
with respect to the
transfer integrals (in the following denoted by the symbol `$set$').\cite{NJP09}

  Then, neglecting for a while the single-ion anisotropy term,
the energy of (classical) spin spiral in the $zx$-plane,
$$
\langle \boldsymbol{S}_i \rangle = S \left( \sin {\bf q}{\bf R}_i, 0, \cos {\bf q}{\bf R}_i \right)
$$
(${\bf R}_i$ being the radius-vector of the site $i$), is given by
$$
E({\bf q}) = -\sum_i \left( J_{0i} \cos {\bf q}{\bf R}_i - d_{0i}^y \sin {\bf q}{\bf R}_i \right),
$$
and the spin-spiral vector ${\bf q}=(q_x,\pi,0)$ in the ground state should correspond
to the minimum of $E({\bf q})$. Obviously, the isotropic exchange interactions
($J_{0i}$) will tend to establish a collinear spin structure with ${\bf q}{\bf R}_i$$=$ $0$ or $\pi$,
while Dzyaloshinskii-Moriya interactions ($d_{0i}^y$) will deform this structure and make
it incommensurate.\cite{Zvezdin}
The Dzyaloshinskii-Moriya interactions are also responsible for the asymmetry between right-handed
($q_x > 0$) and left-handed ($q_x < 0$) spin-spiral states, which is manifested in the
inequality $E({\bf q}) \neq E(-{\bf q})$.

  Parameters of isotropic exchange interactions ($J_{ij}$) are listed in Table~\ref{tab:Heisenberg}.
\begin{table}[h!]
\caption{Isotropic Heisenberg interactions (measured in meV) for PbVO$_3$ and BiCoO$_3$.
Notations of the atomic sites are explained in Fig.~\ref{fig.structure}.
Results of the superexchange theory are denoted by the symbols `$set$'.
Results for infinitesimal spin rotations near the nonrelativistic ground state are
denoted by the symbols `$inf$'.}
\label{tab:Heisenberg}
\begin{ruledtabular}
\begin{tabular}{crrr}
bond & PbVO$_3$ ($set$)      & PbVO$_3$ ($inf$) & BiCoO$_3$ ($inf$) \\
\hline
1-2  & $-49.86$            & $-44.71$       & $-9.65$         \\
1-3  & $-3.63$             & $-0.64$        & $-0.15$         \\
1-4  & $4.76$              & $3.20$         & $-0.91$         \\
1-5  & $3.94$              & $1.25$         & $-1.19$         \\
1-6  & $3.61$              & $1.25$         & $-0.07$         \\
\end{tabular}
\end{ruledtabular}
\end{table}
We note that the schemes `$set$' and `$inf$' in the case of PbVO$_3$ provide very similar results.
This seems to be reasonable, because if the orbital configuration is quenched by the crystal-field
splitting, the spin model (\ref{eqn:SpinModel}) is well defined and the parameters
are not sensitive to the way how they are defined (of course, provided that
$|\hat{t}_{ij}/U| <<1$ and the schemes `$set$' makes a sense).
The magnetic transition temperature, evaluated in the
random-phase approximation (see Ref.~\onlinecite{NJP09} for details) for the G- and C-type
AFM states, is of the order of 200 and 600 K for PbVO$_3$ and
BiCoO$_3$, respectively. The experimental N\'eel temperature for BiCoO$_3$ is 470 K.\cite{Belik06}
The situation in PbVO$_3$ is rather controversial. On the one hand, the
results of the neutron powder diffraction experiment are not conclusive, because their
interpretation strongly depends on the magnetic structure, which was
\textit{assumed} for the analysis of experimental data.\cite{Shpanchenko}
On the other hand, the magnetic susceptibility of PbVO$_3$ does display a broad maximum
at around 200 K, which could be regarded as the sign of an antiferromagnetism. Moreover, the
G-type antiferromagnetic order was proposed for the thin films of PbVO$_3$
below 130 K.\cite{Kumar}

  Parameters of Dzyaloshinskii-Moriya interactions are shown in Table~\ref{tab:DM}.
\begin{table}[h!]
\caption{Nonvanishing
parameters of Dzyaloshinskii-Moriya interactions
${\bf d}_{ij}=(d_{ij}^x,d_{ij}^y,d_{ij}^z)$
(measured in meV) for PbVO$_3$ and BiCoO$_3$.
Notations of atomic sites are explained in Fig.~\ref{fig.structure}.
Other parameters are equal to zero.
Results of the superexchange model are denoted by the symbols `$set$'.
Results for infinitesimal spin rotations near the nonrelativistic ground state are
denoted by the symbols `$inf$'. Note that Dzyaloshinskii-Moriya interactions vanish
in the bond 1-3 due to symmetry constraints.}
\label{tab:DM}
\begin{ruledtabular}
\begin{tabular}{crrr}
parameters           & PbVO$_3$ ($set$) & PbVO$_3$ ($inf$) & BiCoO$_3$ ($inf$) \\
\hline
$d^y_{12}$           & $-0.98$        & $-0.77$        & $-0.13$         \\
$d^y_{14}=-d^x_{14}$ &  $0.39$        &  $0.17$        & $0$             \\
$d^y_{15}$           & $-0.11$        & $-0.04$        & $0$             \\
$d^y_{16}=-d^x_{16}$ & $-0.01$        & $-0.03$        & $0$             \\
\end{tabular}
\end{ruledtabular}
\end{table}
They are at least one order of magnitude smaller than $J_{ij}$ for the same bonds.

  Using these parameters, the spin-spiral vector $q_x$, can be estimated as $q_x a = \pi - \Delta \phi$
($a$ being the lattice parameter in the $xy$-plane), where $\Delta \phi$$=$ $6 \times 10^3 \pi$ and
$4 \times 10^3 \pi$ for PbVO$_3$ and BiCoO$_3$, respectively. Thus,
by considering only $J_{ij}$ and ${\bf d}_{ij}$, both materials are expected to form
spin-spiral structures,
involving more than one hundred unit cells. As we will see below, this scenario indeed holds for PbVO$_3$,
but not for BiCoO$_3$.

  The main difference between PbVO$_3$ and BiCoO$_3$ is in the behavior of the single-ion anisotropy
$\hat{\tau}_{ii}$. For the $S=1/2$ compound PbVO$_3$, $\hat{\tau}_{ii}$ is expected to be zero as the
consequence of fundamental Kramers degeneracy for the odd-electron systems. Particularly, the
ground state of the self-interaction free ion V$^{4+}$ is the Kramers doublet. Therefore, the rotation of spin
corresponds to the unitary transformation of the wave function within this doublet without any
energy cost. The situation is completely different for the $S=2$ (or even-electron) compound
BiCoO$_3$: the Kramers theorem is no longer valid, which formally allows for the finite $\hat{\tau}_{ii}$.
This statement can be verified by direct calculations of the anisotropy energies
$\Delta E = E_\parallel - E_\perp$ (where the symbols ``$\parallel$'' and ``$\perp$''
correspond to the spin configurations, where $\langle \boldsymbol{S}_i \rangle$
is parallel and perpendicular to the tetragonal $z$-axis). In the C-type AFM state,
it yields $\Delta E$$=$ $0.02$ and $-$$5.63$ meV per formula unit for PbVO$_3$ and BiCoO$_3$, respectively.
Moreover, the main contribution to $\Delta E$ indeed originates from the single-ion anisotropy.
This can be seen by repeating the same calculations in the atomic limit (and enforcing $\hat{t}_{ij}$$=$$0$
for all $i$$\neq$$j$), which yields $\Delta E$$=$ $0$ and $-$$5.86$ meV per formula unit for PbVO$_3$ and BiCoO$_3$, respectively.
Small deviations from the atomic limit are due to intersite ($i$$\neq$$j$)
anisotropic interactions $\hat{\tau}_{ij}$,
which can be evaluated in the `$set$'-model and are at least one order of magnitude smaller than ${\bf d}_{ij}$.\cite{NJP09}
Using the obtained values of $\Delta E$ and the symmetry considerations, nonvanishing parameters of the
single-ion anisotropy for BiCoO$_3$ can be estimated as
$\tau_{ii}^{xx} = \tau_{ii}^{yy} = -$$\frac{1}{2} \tau_{ii}^{zz} = 0.49$ meV. Thus, we are dealing the
following hierarchy of magnetic interactions $|J_{ij}|$$>>$$|\hat{\tau}_{ij}|$$>>$$|{\bf d}_{ij}|$.
It means that the formation of the
spin-spiral state in BiCoO$_3$ is strongly affected by the single-ion anisotropies, which will tend to restore the
collinear spin structure by aligning the magnetic moments either parallel or antiparallel to the
$z$-axis. Of course, the final answer about the from of the magnetic ground state of BiCoO$_3$ can be
obtained only on the basis of detailed calculations, which we will discuss in the next section.

\subsection{\label{ElModel} Solution of electronic model}

\begin{figure}[h!]
\begin{center}
\includegraphics[width=16.0cm]{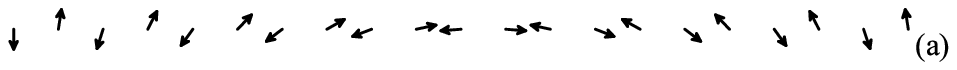}
\includegraphics[width=16.0cm]{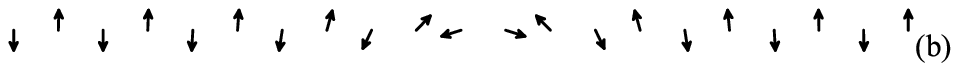}
\end{center}
\caption{\label{fig.SpiralL22}
Spin magnetic structure along the $x$-directions
in the case of PbVO$_3$ (a) and BiCoO$_3$ (b), as obtained in the
Hartree-Fock calculations for $L$$=21$. Here, $x$ is the horizontal axis and
$z$ is the vertical one.}
\end{figure}

\begin{figure}[h!]
\begin{center}
\includegraphics[width=3.5cm]{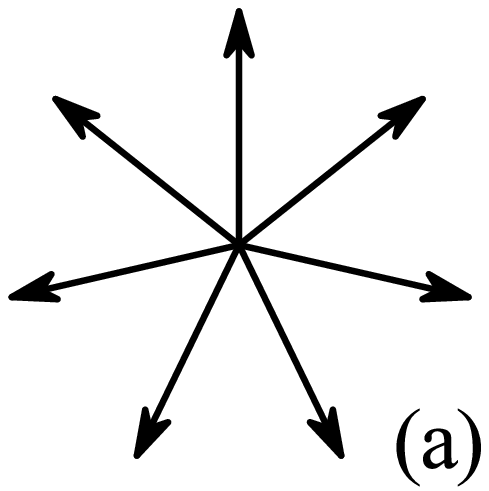}
\includegraphics[width=3.5cm]{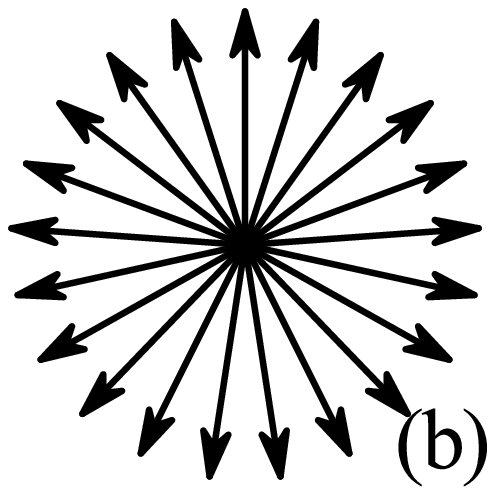}
\includegraphics[width=3.5cm]{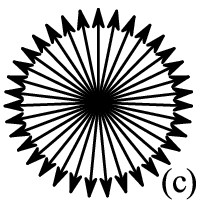}
\includegraphics[width=3.5cm]{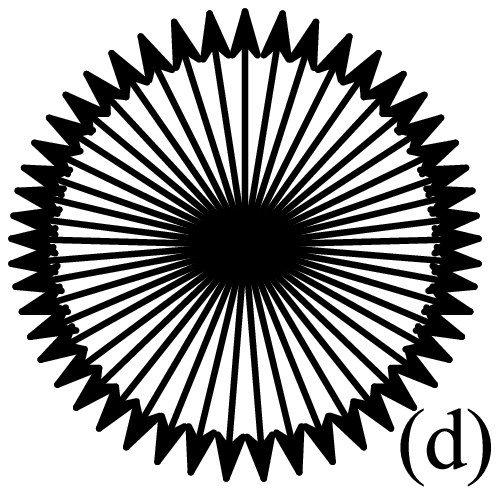}
\end{center}
\begin{center}
\includegraphics[width=3.5cm]{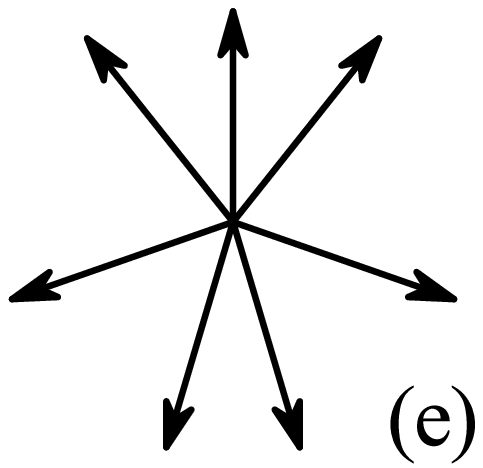}
\includegraphics[width=3.5cm]{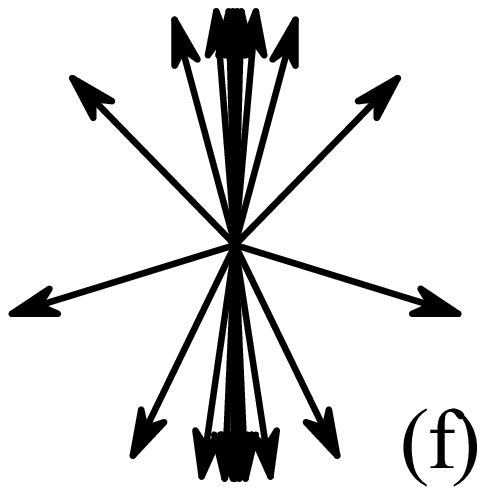}
\includegraphics[width=3.5cm]{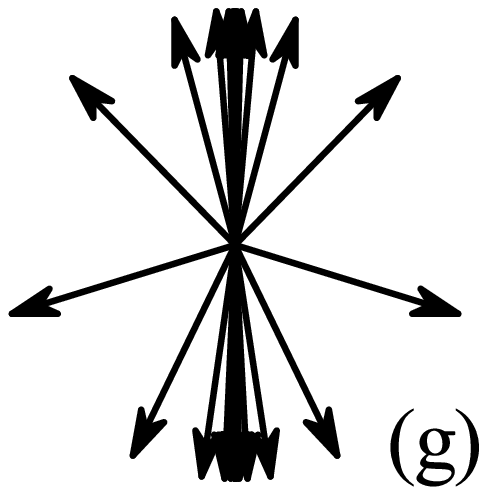}
\includegraphics[width=3.5cm]{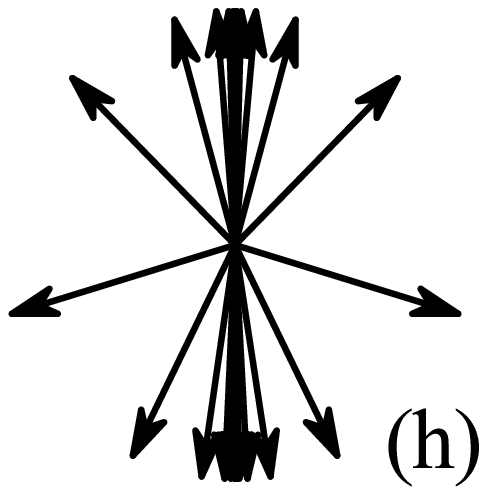}
\end{center}
\caption{\label{fig.SpinDistribution}
Distribution of the spin magnetic moments (as if they were brought to the same origin) in the $xz$-plane
of PbVO$_3$ (top) and BiCoO$_3$ (bottom): results of
Hartree-Fock calculations for $L$$=$ $7$ (`a' and `e'),
$21$ (`b' and `f'), $35$ (`c' and `g'), and $49$ (`d' and `h'). Here, $x$ is the horizontal axis and
$z$ is the vertical one.}
\end{figure}

  In this section we present results of extensive Hartree-Fock calculations for
large supercells, which allow for the spin-spiral solutions with
$q_xa = \pi (|L|-1)/L$, where $|L|$ is the number of cells along the $x$-axis:
$L$ $>0$ and $<0$ corresponds to the right- and left-handed alignment, respectively,
and the limit $|L| \to \infty$ corresponds to the collinear C-type AFM state. The main results
are summarized in Figs.~\ref{fig.SpiralL22} and \ref{fig.SpinDistribution}.
One can clearly see that there is a big difference between PbVO$_3$ and BiCoO$_3$.
PbVO$_3$ tends to form a homogeneous spin-spiral state, where the angle between neighboring magnetic moments
along the $x$-axis remains constant (small deviations are caused by weak inter-site anisotropy
effects). On the contrary, due to the large single-ion anisotropy, the spin-spiral configurations in
BiCoO$_3$ are strongly distorted, and the moments are bunched around the $z$-axis (Figs.~\ref{fig.SpinDistribution}f-h).
The so-called `bunching effect' is well know for magnetic rare-earth metals and was intensively discussed
already more than forty earth ago.\cite{REbunching}
Thus, BiCoO$_3$ tends to form an inhomogeneous magnetic state,
which corresponds to the (nearly) collinear AFM alignment in the wide part of the supercell, except small
`domain wall', where the spins undergo the reorientation within the area of about ten unit cells.
The latter solutions were obtained for odd numbers of cells $L$, which in the AFM lattice results in the formation of the
domain wall defect. For even $L$, the Hartree-Fock equations converge to the C-type AFM state. The spin pattern
in the domain wall is well reproduced already for $L=21$ (Fig.~\ref{fig.SpinDistribution}). For larger cells,
the additional spins participate in the formation of the AFM regions, and are either parallel or antiparallel
to the $z$-axis, leading to tiny changes in Figs.~\ref{fig.SpinDistribution}f-h,
which are practically not distinguishable to the eye.

\begin{figure}[h!]
\begin{center}
\includegraphics[width=8.0cm]{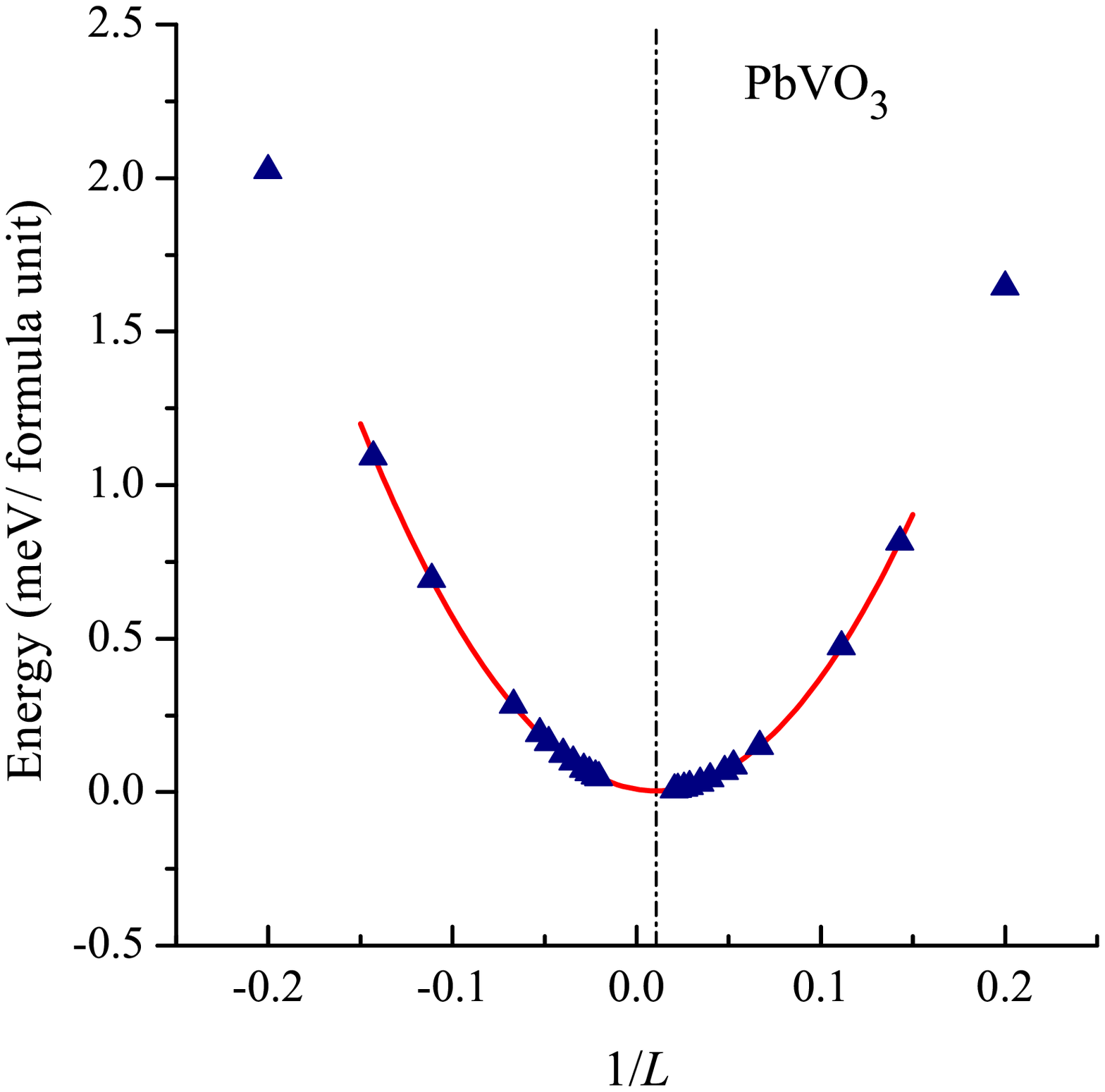}
\includegraphics[width=8.0cm]{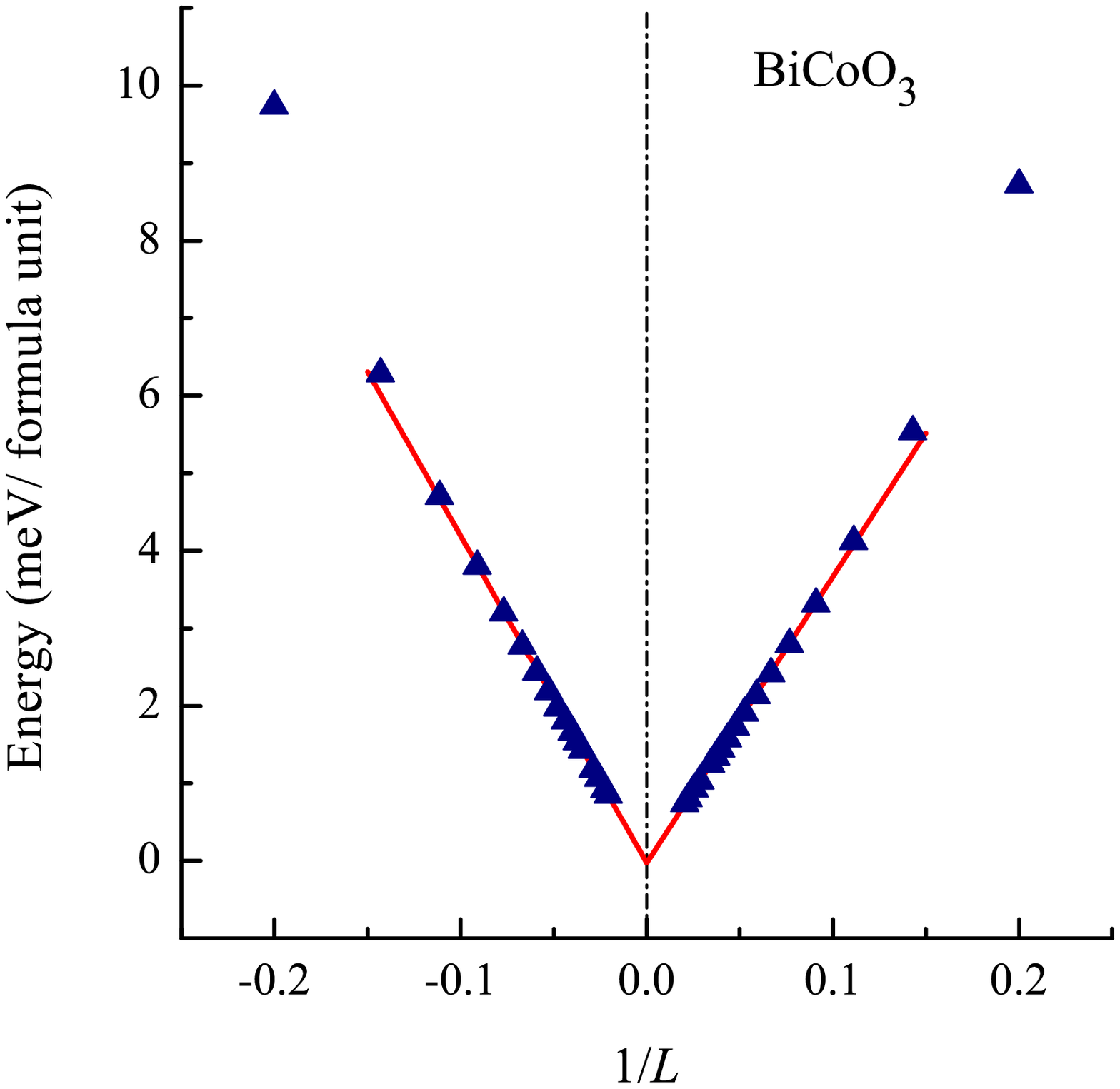}
\end{center}
\caption{\label{fig.TotalEnergy}
(Color online)
Total energies versus $1/L$ as obtained in the
Hartree-Fock approximation for PbVO$_3$ (left) and BiCoO$_3$ (right).
Calculated points are denoted by symbols. Solid line is the results
of interpolation $E=a/L+b/L^2$ in the case of PbVO$_3$
(where $a = -0.99$ meV and $b = 46.36$ meV) and
$E=a/L$ in the case of BiCoO$_3$ (where $a = -42.22$ meV for $L<0$
and $a = 36.94$ meV for $L>0$). The location of the total energy minimum is shown by
the dot-dashed line.}
\end{figure}

  Results of total energy calculations (Fig.~\ref{fig.TotalEnergy}) are well consistent with the above
finding.
As expected for the spin-spiral states, the dependence of the total energy on $1/L$ in the case of
PbVO$_3$ is well described by the parabola. Due to the Dzyaloshinskii-Moriya interactions, there
is a small asymmetry of the total energy with respect to the inversion $L \to -$$L$ of chirality
of the spin spiral. Thus, the total energy minimum, obtained from the extrapolation, corresponds
to the spin-spiral ground state with $L \approx 94$.
On the contrary, the total energy of BiCoO$_3$ is a linear function of $1/L$. This is because of the
localized character of the domain wall, for which the total energy (divided by the total number of
cells) is expected to scale as $1/L$. Thus, the minimum corresponds to the collinear
C-type AFM ground state ($| L | \to \infty$), in which the total energy exhibits the derivative
discontinuity. Nevertheless, even in this case,
the total energy has different slops in the regions $L > 0$ and $L < 0$,
again, due to the difference between the right- and left-handed spin-spiral alignment in the domain wall.

\begin{figure}[h!]
\begin{center}
\includegraphics[width=5cm]{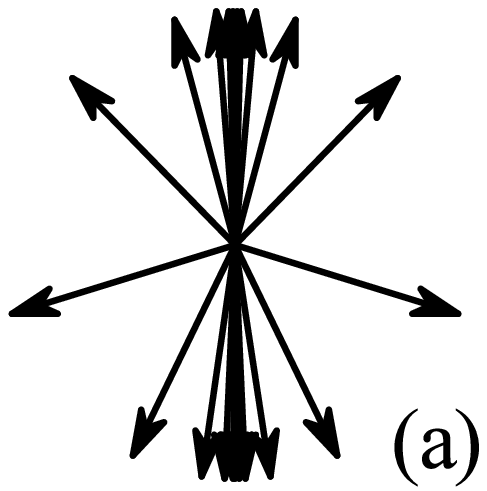} \ \
\includegraphics[width=5cm]{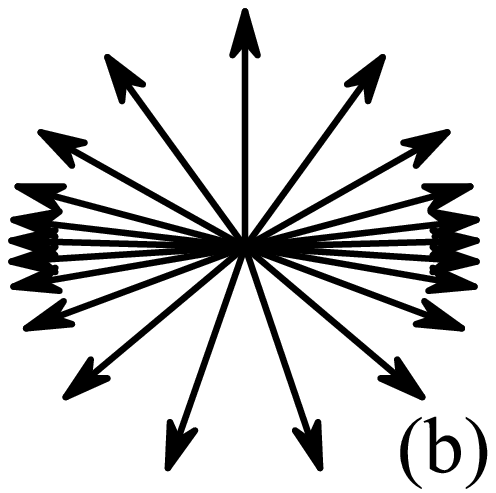}
\end{center}
\caption{\label{fig.BiCoO3spin1}
Distribution of the spin magnetic moments (as if they were brought to the same origin),
obtained in the Hartree-Fock calculations for $L$$=$ $21$.
Left panel (a) shows results for the regular BiCoO$_3$, involving six valence electrons.
Right panel (b) shows the same results for the hypothetical system, which has
the same parameters of electronic Hamiltonian as for BiCoO$_3$
and only one
valence electron.
Thus, according to the Kramers theorem,
the single-ion anisotropy terms should not operate in the case (b).
Here, $x$ is the horizontal axis and
$z$ is the vertical one.}
\end{figure}

  The crucial role of the single-ion anisotropy in the formation of the easy-axis C-type AFM
ground state can be illustrated by repeating supercell calculations for BiCoO$_3$ with the same
parameters of the model Hamiltonian, but with different number of valence electrons:
one instead of six. Thus, according to the Kramers theorem, the single-ion anisotropy should vanish,
similar to PbVO$_3$. The results for $L = 21$ are shown in Fig.~\ref{fig.BiCoO3spin1},
in comparison with regular BiCoO$_3$, including all six valence electrons.
There are two effects. First, as was already discussed in Sec.~\ref{Method},
the effects of SOI are generally larger in BiCoO$_3$, in comparison with PbVO$_3$.
Therefore, inter-site anisotropic interactions become stronger, which is reflected
in some bunching of the spin magnetic moments around the horizontal $x$-axis in the
hypothetical `single-electron BiCoO$_3$' (similar bunching exists in PbVO$_3$ -- Fig.~\ref{fig.SpinDistribution},
but the effect is considerably weaker). Second, the easy-axis alignment in BiCoO$_3$ is solely
related to the single-ion anisotropy term: as long as it is absent in the
hypothetical `single-electron BiCoO$_3$', the spin magnetic moments start to regroup
around the $x$-axis.

\begin{figure}[h!]
\begin{center}
\includegraphics[width=5cm]{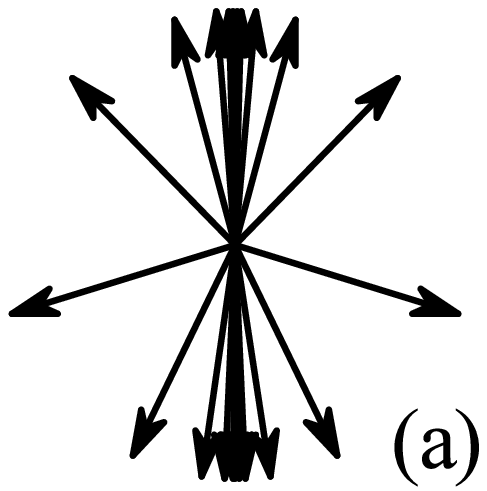} \ \
\includegraphics[width=5cm]{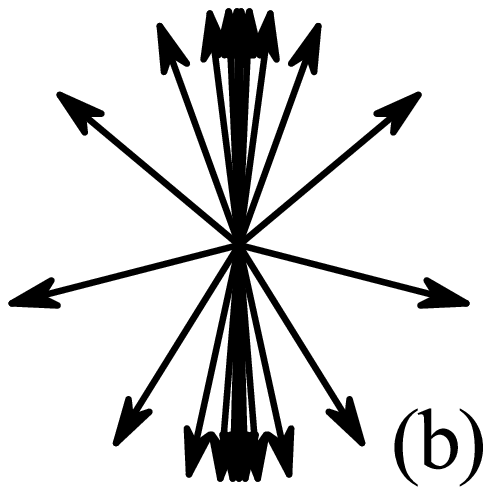}
\end{center}
\caption{\label{fig.BiCoO3spin2}
Influence of the spin-orbit interaction on the magnetic structure of BiCoO$_3$
(results for $L$$=$ $21$). Left panel (a) shows the results, where
the spin-orbit interaction was included only on the Co-sites.
Right panel (b) shows the results, where the spin-orbit interaction was included
on all sites of the system, through the downfolding procedure.
Here, $x$ is the horizontal axis and
$z$ is the vertical one.}
\end{figure}

  Finally, we comment on the dependence of our results on different levels of treatment
of the relativistic SOI. We consider two such schemes: (i) the SOI was included to the
model Hamiltonian as a pseudo-perturbation only at the transition-metal sites,\cite{review2008}
and (ii) the SOI was included at all sites of the system (including heavy Pb and Bi elements)
in the process of downfolding procedure. However, the distribution of the spin magnetic moments,
obtained in these two schemes, is practically indistinguishable (Fig.~\ref{fig.BiCoO3spin2}).
Thus, the SOI at the heavy Pb- and Bi-elements does not seem to play an important role
in the magnetic properties of PbVO$_3$ and BiCoO$_3$.

\section{\label{Summary} Conclusions}

  Being based on results of the low-energy electronic model, derived
from the first-principles electronic structure calculations, we analyzed possible
magnetic structures of two noncentrosymmetric perovskites PbVO$_3$ and BiCoO$_3$.
We have argued that, despite structural similarities, the magnetic behavior of these
two materials is expected to be very different. PbVO$_3$, with the spin $S=1/2$, should form a long-periodic
spin-spiral state, which results solely from the competition between isotropic exchange and
Dzyaloshinskii-Moriya interactions in the noncentrosymmetric crystal structure. Due to the
Kramers degeneracy, the single-ion anisotropy does not operated in PbVO$_3$. However, the latter is expected
to play a major role in BiCoO$_3$, which has the spin $S=2$.
Particularly, the single-ion anisotropy suppress the noncollinear spin-spiral alignment in BiCoO$_3$
and enforces the formation of the C-type antiferromagnetic ground state, in agreement with the experiment.\cite{Belik06}

  We believe that this funding has a direct implication to the properties multiferroic manganites, which also
have spin $S=2$ and the large single-ion anisotropy.\cite{PRL96} Therefore, the numerous claims about the
spin-spiral ground state of these compounds, and related to it improper ferroelectric activity,
should be taken cautiously. Again, due to the large single-ion anisotropy, the ground state of
manganites is not necessary the spin spiral, which prompts a search for alternative mechanism
of multiferroicity in these compounds.\cite{TbMnO3PRB11}

\end{document}